\documentclass{eptcs}
\providecommand{\event}{ADG 2021} 

\usepackage[utf8]{inputenc}
\usepackage{graphicx}
\usepackage{fancyvrb}
\usepackage{color}
\usepackage{mdframed}
\usepackage{argoclp}
\usepackage{amsfonts}
\usepackage{listings}

\usepackage[skins,theorems]{tcolorbox}
\hypersetup{
    colorlinks,
    linkcolor={red!50!black},
    citecolor={blue!50!black},
    urlcolor={blue!80!black}
}

\usepackage{argoclp}

\DefineVerbatimEnvironment%
{CoqVerbatim}{Verbatim}
{frame=lines, fontsize=\small}

\newtheorem{example}{Example}
\newtheorem{theorem}{Theorem}

\def\orcidID#1{\unskip$^{\mbox{\href{https://orcid.org/#1}{\scriptsize{[#1]}} }}$}

\title{Automated Generation of Illustrations for Synthetic Geometry Proofs}

\author{Predrag Jani\v{c}i\'c\orcidID{0000-0001-8922-4948} 
\institute{Department for Computer Science \\ Faculty of Mathematics \\
University of Belgrade \\ Studentski trg 16 \\ 11000 Belgrade, Serbia}
\email{janicic@matf.bg.ac.rs}
\and 
Julien Narboux\orcidID{0000-0003-3527-7184} 
\institute{UMR 7357 CNRS \\ University of Strasbourg \\
P\^{o}le API, Bd S\'ebastien Brant \\ BP 10413 \\ 67412 Illkirch, France}
\email{narboux@unistra.fr}
}

\def\titlerunning{Automated Generation of Illustrations for Synthetic Geometry Proofs}
\def\authorrunning{Janičić \& Narboux}

\begin{document}
\maketitle

\begin{abstract}
We report on a new, simple, modular, and flexible approach for automated 
generation of illustrations for (readable) synthetic geometry proofs. The 
underlying proofs are generated using the Larus automated prover for coherent logic, and 
corresponding illustrations are generated in the GCLC language. Animated 
illustrations are also supported.
\end{abstract}


\section{Introduction}

In geometry, proofs and sketches are closely related.  Humans need sketches to visualize geometry statements, to reason about geometrical problems, and to understand geometry proofs more easily. That is why the use of dynamic geometry software is pervasive in education.
However, many famous examples of incorrect or insufficiently justified steps in geometric reasoning can be attributed to the improper use of visual intuition based on geometric figures.
This goes from unjustified proof steps in Euclid's Elements to proofs and statements  which rely to implicit assumptions from the figure in the context of education \cite{narboux_combining_2021}, passing by incorrect proofs of Euclid's fifth postulate such as the one by Legendre.
Since their early days, automated deduction systems for geometry have also tried to use the knowledge given by the figure\cite{gelernter_empirical_1960}.   
Some authors, also studied how illustrations, following some rules, can provide proofs. This is called diagrammatic reasoning. Diagrammatic reasoning systems can be defined rigorously enough such that they behave as formal systems~\cite{winterstein_dr.doodle:_2004, miller_diagrammatic_2001}. 

In this paper, we present our ongoing work on visual illustrations for geometry proofs.
We are not dealing with diagrammatic reasoning, rather we focus on generation of illustrations for proofs in geometry in the form of sketches similar to what a human would create.  The underlying proofs are readable synthetic proofs generated by our automated theorem prover Larus for coherent logic
\cite{janicic_theorem_2021}.
The use of coherent logic is an essential ingredient of our approach.
The proofs in coherent logic are organized in a purely forward reasoning style. Hence, the fact which has to be proved is not changed during the proof. In logical terms, we use the fact that coherent logic enjoys a complete deduction system which never modifies the right hand sides of the sequents. 
This contributes to the clarity of the visualization.
We illustrate proofs as sketches that get updated by each proof step.
 The initial sketch depicts (a model of) the assumptions. Then, new known facts are introduced 
along the proofs and depicted dynamically by decorating the initial sketch.

    The generated illustrations are stored in a domain specific, geometry oriented, language GCLC (developed by the first 
    author)~\cite{janicic_gclc_2006,janicic_geometry_2010}. The illustrations can also be presented as animations, in a step-by-step manner.

\section{Related Work}

There is a huge number of sources addressing the role of geometric figures in general, so we focus here only on a closely related work about the visualization of proofs in geometry which have either been generated automatically (by automated theorem proving) or machine checked (by interactive theorem proving).

Bertot et al.~\cite{bertot_visualizing_2004} proposed to embed a dynamic geometry software into PCoq~\cite{bertot_generic_1998, amerkad_mathematics_2001} (PCoq is a user interface for the Coq proof assistant~\cite{coq_development_team_the_coq_2010}).
The second author implemented a prototype dynamic geometry software GeoProof which can export statements in the syntax of 
Coq~\cite{narboux_graphical_2007}.
Gao's MMP/geometer software, an automatic geometry theorem prover based on algebraic methods, offered a way to draw figures corresponding to
the statements~\cite{gao_mmp/geometer_2004}.
Wang proposed generating figures from a set of geometrical constraints by decomposing the system of polynomials into irreducible representative triangular sets, and finding an adequate numerical solution from each triangular set~\cite{wang_automated_2003}.
The first author implemented the GCLC\footnote{\url{http://www.matf.bg.ac.rs/~janicic/gclc}}
software in which geometry configurations are described using a custom language, conjectures about configurations
can be expressed and then proved automatically by several available automated theorem provers ~\cite{janicic_gclc_2006, janicic_gclc_2009}.
Pham and Bertot implemented a prototype connecting GeoGebra to PCoq~\cite{pham_combination_2012}. Botana et al.~implemented automated theorem provers based on algebraic methods within GeoGebra
\cite{botana_automated_2015}.

All the above works are focused on the visualization of statements.
Regarding visualization of proofs,
Wilson and Fleuriot designed and implemented a tool for the visualization of proofs as direct acyclic graph with nodes depicted using geometric figures \cite{wilson_combining_2005} for the full-angle method~\cite{chou_automated_1996-1}.
The most significant contribution related to visualization of geometric proofs is the work of Ye et al.~implemented in JGEX~\cite{ye_introduction_2011}.
Ye et al.~use the algebraic characterization and triangulation to generate figures automatically for the statements. JGEX proposes several modes for illustration of proofs: illustration by animated diagrams, or animated diagrams + text, and they apply this method to two kinds of proofs: manually crafted proofs~\cite{ye_visually_2009}, or proofs generated automatically by the full-angle method or the deductive database method~\cite{ye_visually_2010}.


\section{The Context}
\label{sec:context}

\begin{description}
 \item[Synthetic Proofs in Coherent Logic]
A formula of first-order logic is said to be {\em coherent}
if it has the following form:
$$A_0(\vec{x}) \wedge \ldots \wedge A_{n-1}(\vec{x}) \Rightarrow
\exists \vec{y} ( B_0(\vec{x},\vec{y}) \vee \ldots \vee \; B_{m-1}(\vec{x},\vec{y}))$$
\noindent
where universal closure is assumed, and where $0\leq n$, $0\leq m$,
$\vec{x}$ denotes a sequence of variables $x_0, x_2, \ldots, x_{k-1}$ ($0\leq k$), $A_i$ (for $0 \leq i \leq n-1$) denotes an atomic 
formula (involving zero or more variables from $\vec{x}$), 
$\vec{y}$ denotes a sequence of variables 
$y_0, y_2, \ldots, y_{l-1}$ ($0\leq l$), and $B_j$ (for $0 \leq j \leq m-1$)
denotes a conjunction of atomic formulae (involving zero or more 
of the variables from $\vec{x}$ and $\vec{y}$). If $n=0$, then the
left hand side of the implication is assumed to be $\top$ and can 
be omitted. If $m=0$, then the right hand side of the implication 
is assumed to be $\bot$ and can be omitted. There are no function 
symbols with arity greater than zero. Coherent formulae do not 
involve negation. A coherent theory is a set of sentences, 
axiomatised by coherent formulae, and closed under derivability.\footnote{A 
coherent formula is also known as a ``special coherent implication'',
``geometric formula'', ``basic geometric sequent'' 
\cite{dyckhoff_geometrization_2015}. A coherent theory is sometimes called a
``geometric theory''~\cite{maclane_sheaves_1992}. However, a much more widely used notion of ``geometric formula'' allows {\em infinitary disjuctions} 
(but only over finitely many variables)~\cite{vickers_geometric_1993}. Coherent 
formulae involve only finitary disjunctions, so coherent logic 
can be seen as a special case of geometric logic, or as a first-order
fragment of geometric logic.}

A number of theories and theorems can be formulated directly and 
simply in coherent logic (CL). Several authors independently point 
to CL (or rules similar to those of CL) as suitable for expressing (sometimes -- also automating)
significant portions of mathematics~\cite{ganesalingam_fully_2013, avigad_formal_2009}. 
In contrast to resolution-based theorem proving, in CL the conjecture 
being proved is kept unchanged and proved without using refutation, 
Skolemization and clausal form. Thanks to this, CL is suitable for producing readable synthetic proofs~\cite{bezem_automating_2005}.

Every first-order theory has a coherent conservative extension~\cite{dyckhoff_geometrization_2015}, i.e., any first-order theory 
can be translated into CL, possibly with additional predicate 
symbols. 
Translation of FOL formulae into CL involves elimination of 
negations: negations can be kept in place and new predicates 
symbols for corresponding sub-formula have to be introduced, or 
negations can be pushed down to atomic formulae~\cite{polonsky_proofs_2011}. 
In the latter case, for every predicate symbol $R$ (that appears 
in negated form), a new symbol $\overline{R}$ is introduced that 
stands for $\neg R$, and the following axioms are introduced 
$\forall \vec{x} (R(\vec{x}) \wedge \overline{R}(\vec{x}) \Rightarrow \bot)$,
$\forall \vec{x} (R(\vec{x}) \vee \overline{R}(\vec{x}))$.
In order to enable more efficient proving, some advanced translation 
techniques are used. Elimination of function symbols, sometimes called 
{\em anti Skolemization}, is also done by introducing additional 
predicate symbols \cite{nivelle_geometric_2006}.

For a set of coherent axioms $AX$ and the statement 
$A_0(\vec{x}) \wedge \ldots \wedge A_{n-1}(\vec{x}) \Rightarrow
\exists \vec{y} ( B_0(\vec{x},\vec{y}) \vee \ldots \vee \; B_{m-1}(\vec{x},\vec{y}))$
to be proved, within the suitable, very simple CL proof system one has to derive 
$\exists \vec{y} (B_0(\vec{a},\vec{y}) \vee \ldots \vee \; B_{m-1}(\vec{a},\vec{y}))$,
where $\vec{a}$ are fresh constants.
Proofs can be built in the manner of forward reasoning,
as illustrated by the following example.

\begin{example}
\label{ex:basic_ex}
Consider the following set of axioms:

ax1: $\forall x \; ( p(x) \Rightarrow r(x) \vee q(x))$
    
ax2: $\forall x \; ( q(x) \Rightarrow \bot)$

\noindent and the following conjecture that can be proved
as a CL theorem: 

$\forall x \; (p(x) \Rightarrow r(x))$

\begin{tcolorbox}
\newcounter{proofstepnum}
\setcounter{proofstepnum}{0}
\noindent Consider arbitrary $a$ such that: $p(a).$ It should be proved that $r(a)$.
\vspace{5pt}

\proofstep{0}{$r(a)\vee q(a)$ ({\scriptsize by MP, from $p(a)$ using axiom ax1; instantiation:  $X$ $\mapsto$  $a$}) }
\proofstep{1}{Case $r(a)$: }
\proofstep{2}{Proved by assumption! ({\scriptsize by QEDas})}
\proofstep{1}{Case $q(a)$: }
\proofstep{2}{$\bot$ ({\scriptsize by MP, from $q(a)$ using axiom ax2; instantiation:  $X$ $\mapsto$  $a$}) }
\proofstep{2}{Contradiction! ({\scriptsize by QEDefq})}
\proofstep{0}{Proved by case split! ({\scriptsize by QEDcs, by $r(a) , q(a) $})}
\end{tcolorbox}
\end{example}

\item[Larus] is an automated theorem prover for coherent logic. 
It has several underlying proving engines, based on different approaches.
Larus is publicly available and open-source.\footnote{\url{https://github.com/janicicpredrag/Larus}}

\end{description}

\section{Method for Generating Illustrations}
\label{sec:method}

The basic idea of the method is simple: if we know how to visually interpret 
all proof steps that introduce new objects or facts, then we know how to produce a
complete illustration. Indeed, following the Curry-Howard correspondence which relates computer programs and proofs (a proof is a program, and the formula it proves is the type for the program), 
proofs of existential statements can be depicted by a function which takes 
the universally quantified geometric objects of the statement as input and construct 
a witness of the conclusion. The illustration is based on a sequence of
such witnesses in one universum. We will focus on an usual choice --
a Cartesian space (hence, for instance, geometry points will map to 
Cartesian points). Of course, our illustrations will be just 
one model for just one proof branch.

This idea is very well suited to coherent logic as an underlying logical 
framework (see section \ref{sec:context}) and to forward chaining proofs,
which also fits well into CL. In CL proofs, new facts are 
derived using $\mathit{modus ponens}$, so we have to handle only
these rule applications in the proof. 

The above idea, obviously, cannot be applied in conjunction with 
resolution or saturation based theorem provers for first-order 
logic, even less in conjunction with algebraic provers (such as 
those based on Gr\"obner bases or Wu's method), nor it can be 
applied on proof traces produced by such provers. 

\begin{description}
 \item[Rule Applications]
Let us assume that, within the proof, there is an application
of the following axiom (with the obvious intended meaning):
$$\forall x, y \; ( point(x) \wedge point(y) \Rightarrow \exists z \; between(x,z,y))$$
It may have attached the following visual interpretation:
,,for two concrete Cartesian points $a$ and $b$, the Cartesian point 
$c$ such that $c$ is between $a$ and $b$ can be created as the Cartesian
midpoint of $ab$``.

Assuming that the points $a$ and $b$, occurring in the proof are 
associated Cartesian coordinates $(2,5)$ and $(4,11)$, if the above 
axiom has been applied to them, then the new witness point will 
have the associated Cartesian coordinates $(3,8)$. 

Not only  new witnesses can be created, but also
 some new features can also be established and then illustrated. 
In the above example,
we would also draw the segment $ab$. Furthermore, this approach can 
be extended to axioms that do not introduce new objects, but 
only establish some new facts. For instance, if the proof 
establishes that some three points are collinear, then we will
highlight the Cartesian line that contain the three Cartesian points 
corresponding to them.

\item[Axioms vs.~Theorems]
The only thing that we need to illustrate a proof is visual 
interpretation of all applied axioms and lemmas/theorems within
the proof. The visual interpretation of used theorems can 
be produced automatically recursively, using the same approach. Ultimately, 
what we need are only visual interpretations of all axioms,
provided by a human.

\item[Premises and Initial Configuration]
We explained how the figure is updated by each $\mathit{modus ponens}$
rule application. But how do we start the illustration in the
first place? Recall that we prove theorems of the form:
$A_0(\vec{x}) \wedge \ldots \wedge A_{n-1}(\vec{x}) \Rightarrow
\exists \vec{y} ( B_0(\vec{x},\vec{y}) \vee \ldots \vee \; B_{m-1}(\vec{x},\vec{y}))$
(and that all axioms used also have that form). In order to 
build the initial illustration we need some constants $\vec{a}$
such that: 
$A_0(\vec{a}) \wedge \ldots \wedge A_{n-1}(\vec{a})$
holds. But how can we find and illustrate such objects?
We can do that by using the same approach as described above, 
applied to the theorem:
$\exists \vec{x} (A_0(\vec{x}) \wedge \ldots \wedge A_{n-1}(\vec{x})).$\footnote{If, 
instead of just a model of the premises, we want to generate a \emph{construction procedure} for a whole class of initial configurations, we can choose to quantify universally some of the variables, and apply the procedure to the statement:
$\forall \vec{x'} \exists \vec{x''} (A_0(\vec{x'},\vec{x''}) \wedge \ldots \wedge A_{n-1}(\vec{x'},\vec{x''})).$
 Those variables $\vec{x'}$ quantified universally would then correspond to the \emph{free points} of the figure. Depending on the choice of $\vec{x'}$, the statement can be a theorem or not,
 hence, the choice of the set of these free points should be made by the user.}
In fact, here we perform a sort of constraint solving using theorem proving.
If we provide to the prover axioms modeling only ruler-and-compass geometry, then we 
can obtain constructive proof of existence corresponding to a ruler-and-compass construction. 
But, if we provide more powerful axioms to the prover, such as the ability of trisecting the angle, then we could also, in principle, illustrate proofs such as a proof of Morley's theorem (this is only an example because Morley's theorem is currently way out of reach of our prover).
This approach also requires some care: the witness of the existential should be taken as general as possible
(in order to avoid misleading illustration). 
This problem can be addressed by adding non-degenerate conditions, for instance, that 
the points asserted to exist are pairwise distinct and non-collinear.

\item[Case Splits]
There can be many case splits in the proof and it would make no 
sense to illustrate all proof leaves. Indeed, often some
proof branches are contradictory. Not only that they are less
interesting (they typically correspond to degenerative cases)
but they do not have models. 
 Hence, if we have several proof 
branches in one proof node, we do the following:
\begin{itemize}
 \item if all of them end with contradiction, then they all
 belong to some upper contradictory proof branch, and we
 illustrate neither of them;\footnote{If the premises
 themselves are contradictory, then we do not provide
 an illustration. Actually, in some cases we could
 also somehow illustrate contradictory branches (as it 
 is done in some textbooks) -- for instance, if it is proved
 that three points are both colinear and non-colinear, we
 could draw a curved line that connect them. } 
 \item If there are some proof branches that do not end
 with contradiction, then one that corresponds to the model
 being built should be illustrated.
\end{itemize}
We could combine the above with other policies. For instance,
for all case splits of the form $R(\vec{a}) \vee \overline{R}(\vec{a})$,
we could follow only the negative branch, $\overline{R}(\vec{a})$
(for instance, three points are non-collinear) as generally they
correspond to non-degenerated cases.

\item[Randomization]
In order to make illustrations partly unpredictable and
more interesting, some randomization may be added to the
visual interpretation of the axiom. For example, if there
is an axiom stating that for any two distinct points 
there is a third one between them, then in the visual 
counterpart, that third one could be chosen based on
a pseudo-random number between $0$ and~$1$.

\end{description}

\section{Method Implementation}

We implemented the described method within our automated theorem 
prover for coherent logic, Larus \cite{janicic_theorem_2021}. Larus' flexible 
architecture already had proof export to \LaTeX{} and Coq supported. 
So, we have implemented just one class more -- the class 
that exports generated CL proofs (in Larus' internal representation) 
to visual representation. The new code has less than 200 lines of C++.

For the target language, i.e., the language of visual representation
we chose the GCL language~\cite{janicic_geometry_2010}, a rich, special purpose 
language for mathematical, especially geometry illustrations.

For each theorem $\tau$ (given in the TPTP format, the standard format for theorem provers~\cite{sutcliffe_tptp_2009}), once it has been 
proved, the prover generates a file that contains only a function that 
corresponds to that theorem, and a ,,main`` file that includes files 
with all used axioms/theorems, along with the file for $\tau$. 
That ,,main`` file invokes the function for existence of premises for 
$\tau$, and then the function for $\tau$ itself. The prover also 
generates a TPTP formulation of the theorem that corresponds to 
existence of premises for $\tau$ (see Section~\ref{sec:method}). 
Given that theorem was proved, its visual representation can also 
be obtain automatically.

Animations can be obtained simply by showing visualizations
of proof steps one by one or, better still, by showing each 
step in some emphasized manner (or in another color) and
then in a regular way. 

Generated illustrations, stored as readable GCLC files, 
can be further modified and improved by a human.


We present one example in detail in the next section, and
give two more examples in Appendix.

\section{Example: Euclid's {\em Elements}, Book I, Proposition 11}

As a main example, we use Proposition 11 from Euclid's {\em Elements}, 
Book I. Its original form reads as follows: 
,,To draw a straight line at right angles to a given straight line 
from a given point on it.``
The statement represented in first-order logic, following 
the formalization of first book of Euclid's {\em Elements} 
as proposed by Beeson et al.~\cite{beeson_proof-checking_2019},
is the following:

\centerline{
$\forall A, B, C \; \;\mathrm{BetS}\; A C B \; \Rightarrow \; \exists X \;\; \mathrm{Per}\; A C X $ 
}

\noindent
($\mathrm{BetS} \; A C B$ means that $C$ is strictly between $A$ and $B$, 
$\mathrm{Per} \; A C X$ means that $ACX$ is a right angle with the vertex $C$). 
The TPTP file with all needed axioms and lemmas\footnote{The proof 
checked by Coq can be found here:
\url{https://github.com/GeoCoq/GeoCoq/blob/master/Elements/OriginalProofs/proposition_11.v}}
(listed as axioms as well) is as follows:

{\footnotesize
\begin{tcolorbox}
\begin{verbatim}
fof(lemma_betweennotequal,axiom, (! [A,B,C] : 
    ((betS(A,B,C)) => ((( B != C ) & ( A != B ) & ( A != C )))))).
fof(lemma_extension,axiom, (! [A,B,P,Q] : (? [X] : 
    ((( A != B ) & ( P != Q )) => ((betS(A,B,X) & cong(B,X,P,Q))))))).
fof(proposition_01,axiom, (! [A,B] : (? [X] : ((( A != B )) => 
    ((equilateral(A,B,X) & triangle(A,B,X))))))).
fof(defequilateral,axiom, (! [A,B,C] : 
    ((equilateral(A,B,C)) => ((cong(A,B,B,C) & cong(B,C,C,A)))))).
fof(defequilateral2,axiom, (! [A,B,C] : 
    ((cong(A,B,B,C) & cong(B,C,C,A)) => ((equilateral(A,B,C)))))).
fof(lemma_doublereverse,axiom, (! [A,B,C,D] : 
    ((cong(A,B,C,D)) => ((cong(D,C,B,A) & cong(B,A,D,C)))))).
fof(lemma_congruenceflip,axiom, (! [A,B,C,D] : 
    ((cong(A,B,C,D)) => ((cong(B,A,D,C) & cong(B,A,C,D) & cong(A,B,D,C)))))).
fof(defcollinear,axiom, (! [A,B,C] : ((col(A,B,C)) => 
    ((( A = B )) | (( A = C )) | (( B = C )) | 
    (betS(B,A,C)) | (betS(A,B,C)) | (betS(A,C,B)))))).
fof(defcollinear2a,axiom, (! [A,B,C] : ((( A = B )) => ((col(A,B,C)))))).
fof(defcollinear2b,axiom, (! [A,B,C] : ((( A = C )) => ((col(A,B,C)))))).
fof(defcollinear2c,axiom, (! [A,B,C] : ((( B = C )) => ((col(A,B,C)))))).
fof(defcollinear2d,axiom, (! [A,B,C] : ((betS(B,A,C)) => ((col(A,B,C)))))).
fof(defcollinear2e,axiom, (! [A,B,C] : ((betS(A,B,C)) => ((col(A,B,C)))))).
fof(defcollinear2f,axiom, (! [A,B,C] : ((betS(A,C,B)) => ((col(A,B,C)))))).
fof(lemma_collinearorder,axiom, (! [A,B,C] : 
    ((col(A,B,C)) =>  ((col(B,A,C) & col(B,C,A) & col(C,A,B) & col(A,C,B) & col(C,B,A)))))).
fof(deftriangle,axiom, (! [A,B,C] : ((triangle(A,B,C)) => ((~ (col(A,B,C))))))).
fof(deftriangle2,axiom, (! [A,B,C] : ((~(col(A,B,C))) => ((triangle(A,B,C)))))).
fof(defrightangle,axiom, (! [A,B,C] : (? [X] : 
    ((per(A,B,C)) => ((betS(A,B,X) & cong(A,B,X,B) & cong(A,C,X,C) & ( B != C ))))))).
fof(defrightangle2,axiom, (! [A,B,C,X] : 
    ((betS(A,B,X) & cong(A,B,X,B) & cong(A,C,X,C) & ( B != C )) => ((per(A,B,C)))))).
fof(proposition_11,conjecture,(! [A,B,C] : (? [X] : ((betS(A,C,B)) => ((per(A,C,X))))))).
\end{verbatim}
\end{tcolorbox}
}

The proposition was proved automatically by the Larus prover,
giving the following proof in \LaTeX\footnote{
This \LaTeX{} proof presentation is still very verbatim and it provides much more information than a traditional proof; 
for future work we are planning to produce a more natural output.}:

\begin{tcolorbox}
\begin{theorem}
proposition\_11 : $\forall A \; \forall B \; \forall C \; ( betS(A, C, B) \Rightarrow \exists X \; (per(A, C, X))\;)$
\end{theorem}

\setcounter{proofstepnum}{0}

\noindent{\em Proof:}\\
\vspace{5pt}
\hspace{0.7cm}\begin{minipage}{0.95\textwidth}
\noindent Consider arbitrary $a$, $b$, $c$ such that:  $betS(a, c, b)$. It should be proved that $\exists X\; \; per(a, c, X)$.
\vspace{5pt}

\proofstep{0}{Let $w$ be such that $betS(a, c, w)\wedge cong(c, w, a, c)$ ({\scriptsize by MP, from $betS(a, c, b)$, $betS(a, c, b)$ using axiom lemma\_extension; instantiation:  $A$ $\mapsto$  $a$,  $B$ $\mapsto$  $c$,  $P$ $\mapsto$  $a$,  $Q$ $\mapsto$  $c$}) }
\proofstep{0}{Let $w1$ be such that $equilateral(a, w, w1)\wedge triangle(a, w, w1)$ ({\scriptsize by MP, from $betS(a, c, w)\wedge cong(c, w, a, c)$ using axiom proposition\_01; instantiation:  $A$ $\mapsto$  $a$,  $B$ $\mapsto$  $w$}) }
\proofstep{0}{$w1 = c\vee w1 \neq c$ ({\scriptsize by MP, using axiom eq\_excluded\_middle; instantiation:  $A$ $\mapsto$  $w1$,  $B$ $\mapsto$  $c$}) }
\proofstep{1}{Case $w1 = c$: }
\proofstep{2}{$col(a, w, w1)$ ({\scriptsize by MP, from $betS(a, c, w)\wedge cong(c, w, a, c)$, $w1 = c$ using axiom colEqSub2; instantiation:  $A$ $\mapsto$  $a$,  $B$ $\mapsto$  $w$,  $C$ $\mapsto$  $c$,  $X$ $\mapsto$  $w1$}) }
\proofstep{2}{$\bot$ ({\scriptsize by MP, from $col(a, w, w1)$, $equilateral(a, w, w1)\wedge triangle(a, w, w1)$ using axiom nnncolNegElim; instantiation:  $A$ $\mapsto$  $a$,  $B$ $\mapsto$  $w$,  $C$ $\mapsto$  $w1$}) }
\proofstep{2}{Contradiction! ({\scriptsize by QEDefq})}
\proofstep{1}{Case $w1 \neq c$: }
\proofstep{2}{$per(a, c, w1)$ ({\scriptsize by MP, from $betS(a, c, w)\wedge cong(c, w, a, c)$, $betS(a, c, w)\wedge cong(c, w, a, c)$, $equilateral(a, w, w1)\wedge triangle(a, w, w1)$, $w1 \neq c$ using axiom defrightangle2; instantiation:  $A$ $\mapsto$  $a$,  $B$ $\mapsto$  $c$,  $C$ $\mapsto$  $w1$,  $X$ $\mapsto$  $w$}) }
\proofstep{2}{Proved by assumption! ({\scriptsize by QEDas})}
\proofstep{0}{Proved by case split! ({\scriptsize by QEDcs, by $w1 = c , w1 \neq c $})}
\end{minipage}
\end{tcolorbox}

Note that Larus' proofs omits applications of ,,simple axioms``, 
axioms that are universal implications from one atomic formula to 
another. The above proof explicitly uses only the 
following lemmas (not counting those implied by equality axioms and 
those introducing $\bot$): {\em lemma\_extension}, {\em proposition\_01}, 
{\em defrightangle2}. Hence, we need visual interpretation of these. 
In order to obtain them, we can run the prover or provide the visual 
interpretation ourselves. For instance, the visual counterpart for the 
famous Euclid's proposition 1 (on existence of a equilateral triangle on 
a given segment), can be like the following:

\begin{tcolorbox}
{\footnotesize 
\begin{verbatim}
procedure proposition_01 { A B X } { 
  circle c1 A B  
  drawcircle c1
  circle c2 B A
  drawcircle c2
  intersec2 X2 X c1 c2
  cmark X 
} 
\end{verbatim}
}
\end{tcolorbox}

\noindent
In this visual interpretation, the point $X$ is obtained (in the 
Cartesian model) as an intersection of two circles -- one with
the center $A$ with $B$ on it, and one the center $B$ with $A$
on it (the 
command \verb|cmark X| annotates the point $X$ by a small circle). 
Of course, one could have chosen the other intersection 
of two circles.
 Again, instead of providing this 
visual interpretation, we could run the prover on {\em proposition\_01} 
and could get another function, expressed in terms of axioms 
(or, possibly, some other lemmas).

The prover also generates the conjecture which establishes existence 
of objects such that the premises of the main conjecture hold:

\begin{tcolorbox}
{\footnotesize 
\begin{verbatim}
fof(proposition_11, axiom, ( ? [A,B,C] : (betS(A,C,B)))) 
\end{verbatim}
}
\end{tcolorbox}

\noindent 
This conjecture can be proved  
automatically or can be illustrated by hand, as here (within the file 
\verb|proposition_11_exists.gcl|):

\begin{tcolorbox}
{\footnotesize 
\begin{verbatim}
procedure proposition_11_exists { a b c  }  { 
  point a 8 2 
  point b 22 7
  towards c a b 0.7
  cmark_t a
  cmark_t b
  cmark_t c
}
\end{verbatim}
}
\end{tcolorbox}

The main conjecture is described by the following function, generated automatically by the prover
(within the file \verb|proposition_11.gcl|):

\begin{tcolorbox}
{\footnotesize 
\begin{verbatim}
procedure proposition_11 { a b c w  }  { 
  call lemma_extension { a c a c w } 
  mark_t w
  call proposition_01 { a w w1 } 
  mark_t w1
  % --- Illustration for branch 2
  call defrightangle2 { a c w1 w } 
} 
\end{verbatim}
}
\end{tcolorbox}
\noindent 
One point ($w$) is introduced by the application of {\em lemma\_extension}, 
and another ($w1$) by the application of {\em proposition\_01}. 
As we can see, the illustration follows the second, non-contradictory
branch of the proof. The final proof step does not introduce
new objects, but only establishes a property $ac \perp cw_1$
(which is stressed by drawing a small square at the vertex
of the angle).

Finally, the main file, that will be processed by GCLC, is 
generated automatically and looks like the following:

\begin{tcolorbox}
{\footnotesize 
\begin{verbatim}
% ----- Proof illustration -----
include proposition_11.gcl
include proposition_11_exists.gcl
include lemma_extension.gcl
include proposition_01.gcl
include defrightangle2.gcl
%-----------------------------
call proposition_11_exists { a b c } 
call proposition_11 { a b c w  } 
\end{verbatim}
}
\end{tcolorbox}

After the ,,include`` section with all needed files/functions,
there is a call to the function which illustrates the objects
such that the premises hold. After that, with the concrete
points \verb|a|, \verb|b|, \verb|c| (provided by the previous
function), 
the main function is invoked. 
The illustration can be modified such that it provides animation
(within GCLC graphical environment, but it can be also 
exported as a sequence of images). A visual counterpart of 
each step is shown first in red, then in black and it remains 
like that. This is implemented using layers, as in the
following addition to the main GCLC file, which ensures that 
all levels from $i$ ($i=1, 2, 3, \ldots$) are hidden (in our 
example, there are 6 layers):

\begin{tcolorbox}
{\footnotesize 
\begin{verbatim}
animation_frames 7 1
point A0 0 0 
point A1 1 0 7 0  
distance dA A0 A1 
hide_layers_from dA 
\end{verbatim}
}
\end{tcolorbox}

\begin{figure}
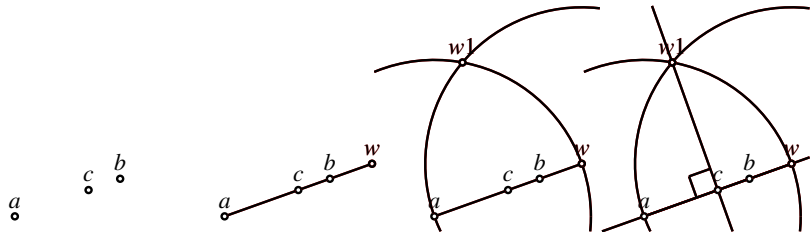

\begin{center}
\hspace*{-9mm}
\input{PROOF025_proposition_11pQF_BV_illustration-1.tkz}  
\hspace*{-5mm}
\input{PROOF025_proposition_11pQF_BV_illustration-2.tkz}  
\hspace*{-5mm}
\input{PROOF025_proposition_11pQF_BV_illustration-3.tkz}  
\hspace*{-5mm}
\input{PROOF025_proposition_11pQF_BV_illustration-4.tkz}  
\caption{Four steps in illustration of the proposition 11}
\label{fig:prop11} 
\end{center}
\end{figure}

Processing the above GCLC file gives the animation, 
i.e., the illustration that gets extended in several steps,
as shown in Figure \ref{fig:prop11}. Within GCLC environment, 
each step is shown first in red, then in black colour.
(Figures can be exported to different formats, e.g.~
\verb|TikZ|.)

\section{Conclusions and Perspectives}

We presented a new approach for automated generation of illustrations of proofs 
of geometry theorems. The approach is simple, as it is a small
extension to our prover for coherent logic, Larus. The approach
is modular, as all illustrations rely only on visual interpretations
of axioms used. The approach is flexible, as one can provide
by hand different visual counterparts of the axioms, but also
of particular lemmas used within other proofs. 
The illustrations are generated in the form of the GCL language and
they can also be viewed as animations, where proof steps
are shown step-by-step.

When Larus finds a proof, it can generate the corresponding Coq proof and illustration. 
In the future,  this work could be extended to deal with coherent logic proofs which are not necessarily obtained automatically by Larus, but constructed interactively within Coq.

We also plan to improve the translation of the proof to text form, 
to generate a more natural English text.

\bibliography{biblio}

\begin{thebibliography}{10}
\providecommand{\bibitemdeclare}[2]{}
\providecommand{\surnamestart}{}
\providecommand{\surnameend}{}
\providecommand{\urlprefix}{Available at }
\providecommand{\url}[1]{\texttt{#1}}
\providecommand{\href}[2]{\texttt{#2}}
\providecommand{\urlalt}[2]{\href{#1}{#2}}
\providecommand{\doi}[1]{doi:\urlalt{http://dx.doi.org/#1}{#1}}
\providecommand{\eprint}[1]{arXiv:\urlalt{https://arxiv.org/abs/#1}{#1}}
\providecommand{\bibinfo}[2]{#2}

\bibitemdeclare{inproceedings}{amerkad_mathematics_2001}
\bibitem{amerkad_mathematics_2001}
\bibinfo{author}{Ahmed \surnamestart Amerkad\surnameend}, \bibinfo{author}{Yves
  \surnamestart Bertot\surnameend}, \bibinfo{author}{Loïc \surnamestart
  Pottier\surnameend} \& \bibinfo{author}{Laurence \surnamestart
  Rideau\surnameend} (\bibinfo{year}{2001}): \emph{\bibinfo{title}{Mathematics
  and {Proof} {Presentation} in {Pcoq}}}.
\newblock In: {\sl \bibinfo{booktitle}{Workshop {Proof} {Transformation} and
  {Presentation} and {Proof} {Complexities} in connection with {IJCAR} 2001}},
  \bibinfo{address}{Siena}.

\bibitemdeclare{article}{avigad_formal_2009}
\bibitem{avigad_formal_2009}
\bibinfo{author}{Jeremy \surnamestart Avigad\surnameend},
  \bibinfo{author}{Edward \surnamestart Dean\surnameend} \&
  \bibinfo{author}{John \surnamestart Mumma\surnameend} (\bibinfo{year}{2009}):
  \emph{\bibinfo{title}{A {Formal} {System} for {Euclid}'s {Elements}}}.
\newblock {\sl \bibinfo{journal}{The Review of Symbolic Logic}}
  \bibinfo{volume}{2}, pp. \bibinfo{pages}{700--768},
  \doi{10.1017/S1755020309990098}.

\bibitemdeclare{article}{beeson_proof-checking_2019}
\bibitem{beeson_proof-checking_2019}
\bibinfo{author}{Michael \surnamestart Beeson\surnameend},
  \bibinfo{author}{Julien \surnamestart Narboux\surnameend} \&
  \bibinfo{author}{Freek \surnamestart Wiedijk\surnameend}
  (\bibinfo{year}{2019}): \emph{\bibinfo{title}{Proof-checking {Euclid}}}.
\newblock {\sl \bibinfo{journal}{Annals of Mathematics and Artificial
  Intelligence}} \bibinfo{volume}{85}(\bibinfo{number}{2-4}), pp.
  \bibinfo{pages}{213--257}, \doi{10.1007/s10472-018-9606-x}.
\newblock \bibinfo{note}{Publisher: Springer}.

\bibitemdeclare{article}{bertot_visualizing_2004}
\bibitem{bertot_visualizing_2004}
\bibinfo{author}{Yves \surnamestart Bertot\surnameend},
  \bibinfo{author}{Frédérique \surnamestart Guilhot\surnameend} \&
  \bibinfo{author}{Loïc \surnamestart Pottier\surnameend}
  (\bibinfo{year}{2004}): \emph{\bibinfo{title}{Visualizing {Geometrical}
  {Statements} with {GeoView}}}.
\newblock {\sl \bibinfo{journal}{Proceedings of the Workshop User Interfaces
  for Theorem Provers 2003}} \bibinfo{volume}{103}, pp.
  \bibinfo{pages}{49--65}, \doi{10.1016/j.entcs.2004.09.013}.

\bibitemdeclare{article}{bertot_generic_1998}
\bibitem{bertot_generic_1998}
\bibinfo{author}{Yves \surnamestart Bertot\surnameend} \&
  \bibinfo{author}{Laurent \surnamestart Thery\surnameend}
  (\bibinfo{year}{1998}): \emph{\bibinfo{title}{A {Generic} {Approach} to
  {Building} {User} {Interfaces} for {Theorem} {Provers}}}.
\newblock {\sl \bibinfo{journal}{The Journal of Symbolic Computation}}
  \bibinfo{volume}{25}, pp. \bibinfo{pages}{161--194},
  \doi{10.1006/jsco.1997.0171}.

\bibitemdeclare{inproceedings}{bezem_automating_2005}
\bibitem{bezem_automating_2005}
\bibinfo{author}{Marc \surnamestart Bezem\surnameend} \&
  \bibinfo{author}{Thierry \surnamestart Coquand\surnameend}
  (\bibinfo{year}{2005}): \emph{\bibinfo{title}{Automating {Coherent}
  {Logic}}}.
\newblock In \bibinfo{editor}{Geoff \surnamestart Sutcliffe\surnameend} \&
  \bibinfo{editor}{Andrei \surnamestart Voronkov\surnameend}, editors: {\sl
  \bibinfo{booktitle}{12th {International} {Conference} on {Logic} for
  {Programming}, {Artificial} {Intelligence}, and {Reasoning} — {LPAR}
  2005}}, {\sl \bibinfo{series}{Lecture {Notes} in {Computer} {Science}}}
  \bibinfo{volume}{3835}, \bibinfo{publisher}{Springer-Verlag}, pp.
  \bibinfo{pages}{246--260}, \doi{10.1007/11591191\_18}.

\bibitemdeclare{article}{botana_automated_2015}
\bibitem{botana_automated_2015}
\bibinfo{author}{Francisco \surnamestart Botana\surnameend},
  \bibinfo{author}{Markus \surnamestart Hohenwarter\surnameend},
  \bibinfo{author}{Predrag \surnamestart Janičić\surnameend},
  \bibinfo{author}{Zoltán \surnamestart Kovács\surnameend},
  \bibinfo{author}{Ivan \surnamestart Petrović\surnameend},
  \bibinfo{author}{Tomás \surnamestart Recio\surnameend} \&
  \bibinfo{author}{Simon \surnamestart Weitzhofer\surnameend}
  (\bibinfo{year}{2015}): \emph{\bibinfo{title}{Automated {Theorem} {Proving}
  in {GeoGebra}: {Current} {Achievements}}}.
\newblock {\sl \bibinfo{journal}{Journal of Automated Reasoning}}
  \bibinfo{volume}{55}(\bibinfo{number}{1}), pp. \bibinfo{pages}{39--59},
  \doi{10.1007/s10817-015-9326-4}.

\bibitemdeclare{article}{chou_automated_1996-1}
\bibitem{chou_automated_1996-1}
\bibinfo{author}{Shang-Ching \surnamestart Chou\surnameend},
  \bibinfo{author}{Xiao-Shan \surnamestart Gao\surnameend} \&
  \bibinfo{author}{Ji~\surnamestart Zhang\surnameend} (\bibinfo{year}{1996}):
  \emph{\bibinfo{title}{Automated {Generation} of {Readable} {Proofs} with
  {Geometric} {Invariants}, {II}. {Theorem} {Proving} {With} {Full}-{Angles}}}.
\newblock {\sl \bibinfo{journal}{Journal of Automated Reasoning}}
  \bibinfo{volume}{17}(\bibinfo{number}{13}), pp. \bibinfo{pages}{349--370},
  \doi{10.1007/BF00283134}.

\bibitemdeclare{book}{coq_development_team_the_coq_2010}
\bibitem{coq_development_team_the_coq_2010}
\bibinfo{author}{\surnamestart {Coq development team, The}\surnameend}
  (\bibinfo{year}{2010}): \emph{\bibinfo{title}{The {Coq} proof assistant
  reference manual, {Version} 8.3}}.
\newblock \bibinfo{publisher}{LogiCal Project}.
\newblock \urlprefix\url{http://coq.inria.fr}.

\bibitemdeclare{article}{dyckhoff_geometrization_2015}
\bibitem{dyckhoff_geometrization_2015}
\bibinfo{author}{Roy \surnamestart Dyckhoff\surnameend} \&
  \bibinfo{author}{Sara \surnamestart Negri\surnameend} (\bibinfo{year}{2015}):
  \emph{\bibinfo{title}{Geometrization of first-order logic}}.
\newblock {\sl \bibinfo{journal}{The Bulletin of Symbolic Logic}}
  \bibinfo{volume}{21}, pp. \bibinfo{pages}{123--163},
  \doi{10.1017/bsl.2015.7}.

\bibitemdeclare{article}{ganesalingam_fully_2013}
\bibitem{ganesalingam_fully_2013}
\bibinfo{author}{M.~\surnamestart Ganesalingam\surnameend} \&
  \bibinfo{author}{W.~T. \surnamestart Gowers\surnameend}
  (\bibinfo{year}{2013}): \emph{\bibinfo{title}{A fully automatic problem
  solver with human-style output}}.
\newblock {\sl \bibinfo{journal}{CoRR}} \bibinfo{volume}{abs/1309.4501}.

\bibitemdeclare{inproceedings}{gao_mmp/geometer_2004}
\bibitem{gao_mmp/geometer_2004}
\bibinfo{author}{Xiao-Shan \surnamestart Gao\surnameend} \&
  \bibinfo{author}{Qiang \surnamestart Lin\surnameend} (\bibinfo{year}{2004}):
  \emph{\bibinfo{title}{{MMP}/{Geometer} - {A} {Software} {Package} for
  {Automated} {Geometric} {Reasoning}}}.
\newblock In: {\sl \bibinfo{booktitle}{Proceedings of {Automated} {Deduction}
  in {Geometry} ({ADG02})}}, {\sl \bibinfo{series}{Lecture {Notes} in
  {Computer} {Science}}} \bibinfo{volume}{2930},
  \bibinfo{publisher}{Springer-Verlag}, pp. \bibinfo{pages}{44--66},
  \doi{10.1007/978-3-540-24616-9\_4}.

\bibitemdeclare{inproceedings}{gelernter_empirical_1960}
\bibitem{gelernter_empirical_1960}
\bibinfo{author}{Herbert \surnamestart Gelernter\surnameend},
  \bibinfo{author}{J.~R. \surnamestart Hansen\surnameend} \&
  \bibinfo{author}{Donald \surnamestart Loveland\surnameend}
  (\bibinfo{year}{1960}): \emph{\bibinfo{title}{Empirical explorations of the
  geometry theorem machine}}.
\newblock In: {\sl \bibinfo{booktitle}{Papers presented at the {May} 3-5, 1960,
  western joint {IRE}-{AIEE}-{ACM} computer conference}},
  \bibinfo{series}{{IRE}-{AIEE}-{ACM} '60 ({Western})},
  \bibinfo{publisher}{ACM}, \bibinfo{address}{San Francisco, California}, pp.
  \bibinfo{pages}{143--149}, \doi{10.1145/1460361.1460381}.

\bibitemdeclare{incollection}{janicic_gclc_2006}
\bibitem{janicic_gclc_2006}
\bibinfo{author}{Predrag \surnamestart Janičić\surnameend}
  (\bibinfo{year}{2006}): \emph{\bibinfo{title}{{GCLC} — {A} {Tool} for
  {Constructive} {Euclidean} {Geometry} and {More} {Than} {That}}}.
\newblock In \bibinfo{editor}{Andrés \surnamestart Iglesias\surnameend} \&
  \bibinfo{editor}{Nobuki \surnamestart Takayama\surnameend}, editors: {\sl
  \bibinfo{booktitle}{Mathematical {Software} - {ICMS} 2006}}, {\sl
  \bibinfo{series}{Lecture {Notes} in {Computer} {Science}}}
  \bibinfo{volume}{4151}, \bibinfo{publisher}{Springer}, pp.
  \bibinfo{pages}{58--73}, \doi{10.1007/11832225\_6}.

\bibitemdeclare{unpublished}{janicic_gclc_2009}
\bibitem{janicic_gclc_2009}
\bibinfo{author}{Predrag \surnamestart Janičić\surnameend}
  (\bibinfo{year}{2009}): \emph{\bibinfo{title}{{GCLC} 9.0/{WinGCLC} 2009}}.
\newblock \bibinfo{note}{Manual for the GCLC Dynamic Geometry Software}.

\bibitemdeclare{article}{janicic_geometry_2010}
\bibitem{janicic_geometry_2010}
\bibinfo{author}{Predrag \surnamestart Janičić\surnameend}
  (\bibinfo{year}{2010}): \emph{\bibinfo{title}{Geometry {Constructions}
  {Language}}}.
\newblock {\sl \bibinfo{journal}{Journal of Automated Reasoning}}
  \bibinfo{volume}{44}(\bibinfo{number}{1-2}), pp. \bibinfo{pages}{3--24},
  \doi{10.1007/s10817-009-9135-8}.

\bibitemdeclare{misc}{janicic_theorem_2021}
\bibitem{janicic_theorem_2021}
\bibinfo{author}{Predrag \surnamestart Janičić\surnameend} \&
  \bibinfo{author}{Julien \surnamestart Narboux\surnameend}
  (\bibinfo{year}{2021}): \emph{\bibinfo{title}{Theorem {Proving} as
  {Constraint} {Solving} with {Coherent} {Logic}}}.
\newblock \bibinfo{note}{Submitted}.

\bibitemdeclare{book}{maclane_sheaves_1992}
\bibitem{maclane_sheaves_1992}
\bibinfo{author}{Saunders \surnamestart MacLane\surnameend} \&
  \bibinfo{author}{Ieke \surnamestart Moerdijk\surnameend}
  (\bibinfo{year}{1992}): \emph{\bibinfo{title}{Sheaves in geometry and logic:
  a first introduction to topos theory}}.
\newblock \bibinfo{publisher}{Springer-Verlag}.

\bibitemdeclare{phdthesis}{miller_diagrammatic_2001}
\bibitem{miller_diagrammatic_2001}
\bibinfo{author}{Nathaniel \surnamestart Miller\surnameend}
  (\bibinfo{year}{2001}): \emph{\bibinfo{title}{A diagrammatic formal system
  for {Euclidean} geometry}}.
\newblock Ph.D. thesis, \bibinfo{school}{Cornell University}.

\bibitemdeclare{article}{narboux_graphical_2007}
\bibitem{narboux_graphical_2007}
\bibinfo{author}{Julien \surnamestart Narboux\surnameend}
  (\bibinfo{year}{2007}): \emph{\bibinfo{title}{A {Graphical} {User}
  {Interface} for {Formal} {Proofs} in {Geometry}}}.
\newblock {\sl \bibinfo{journal}{Journal of Automated Reasoning}}
  \bibinfo{volume}{39}(\bibinfo{number}{2}), pp. \bibinfo{pages}{161--180},
  \doi{10.1007/s10817-007-9071-4}.

\bibitemdeclare{incollection}{narboux_combining_2021}
\bibitem{narboux_combining_2021}
\bibinfo{author}{Julien \surnamestart Narboux\surnameend} \&
  \bibinfo{author}{Viviane \surnamestart Durand-Guerrier\surnameend}
  (\bibinfo{year}{2021}): \emph{\bibinfo{title}{Combining pencil/paper proofs
  and formal proofs, a challenge for {Artificial} {Intelligence} and
  mathematics education}}.
\newblock In: {\sl \bibinfo{booktitle}{Mathematics {Education} in the {Age} of
  {Artificial} {Intelligence}: {How} {Intelligence} can serve mathematical
  human learning}}, \bibinfo{publisher}{Springer}.
\newblock \bibinfo{note}{In press}.

\bibitemdeclare{inproceedings}{nivelle_geometric_2006}
\bibitem{nivelle_geometric_2006}
\bibinfo{author}{Hans~de \surnamestart Nivelle\surnameend} \&
  \bibinfo{author}{Jia \surnamestart Meng\surnameend} (\bibinfo{year}{2006}):
  \emph{\bibinfo{title}{Geometric {Resolution}: {A} {Proof} {Procedure} {Based}
  on {Finite} {Model} {Search}}}.
\newblock In \bibinfo{editor}{Ulrich \surnamestart Furbach\surnameend} \&
  \bibinfo{editor}{Natarajan \surnamestart Shankar\surnameend}, editors: {\sl
  \bibinfo{booktitle}{Automated {Reasoning}, {Third} {International} {Joint}
  {Conference}, {IJCAR} 2006, {Seattle}, {WA}, {USA}, {August} 17-20, 2006,
  {Proceedings}}}, {\sl \bibinfo{series}{Lecture {Notes} in {Computer}
  {Science}}} \bibinfo{volume}{4130}, \bibinfo{publisher}{Springer}, pp.
  \bibinfo{pages}{303--317}, \doi{10.1007/11814771\_28}.

\bibitemdeclare{article}{pham_combination_2012}
\bibitem{pham_combination_2012}
\bibinfo{author}{Tuan~Minh \surnamestart Pham\surnameend} \&
  \bibinfo{author}{Yves \surnamestart Bertot\surnameend}
  (\bibinfo{year}{2012}): \emph{\bibinfo{title}{A {Combination} of a {Dynamic}
  {Geometry} {Software} {With} a {Proof} {Assistant} for {Interactive} {Formal}
  {Proofs}}}.
\newblock {\sl \bibinfo{journal}{Electron. Notes Theor. Comput. Sci.}}
  \bibinfo{volume}{285}, pp. \bibinfo{pages}{43--55},
  \doi{10.1016/j.entcs.2012.06.005}.

\bibitemdeclare{phdthesis}{polonsky_proofs_2011}
\bibitem{polonsky_proofs_2011}
\bibinfo{author}{Andrew \surnamestart Polonsky\surnameend}
  (\bibinfo{year}{2011}): \emph{\bibinfo{title}{Proofs, {Types} and {Lambda}
  {Calculus}}}.
\newblock Ph.D. thesis, \bibinfo{school}{University of Bergen}.

\bibitemdeclare{article}{sutcliffe_tptp_2009}
\bibitem{sutcliffe_tptp_2009}
\bibinfo{author}{G.~\surnamestart Sutcliffe\surnameend} (\bibinfo{year}{2009}):
  \emph{\bibinfo{title}{The {TPTP} {Problem} {Library} and {Associated}
  {Infrastructure}: {The} {FOF} and {CNF} {Parts}, v3.5.0}}.
\newblock {\sl \bibinfo{journal}{Journal of Automated Reasoning}}
  \bibinfo{volume}{43}(\bibinfo{number}{4}), pp. \bibinfo{pages}{337--362},
  \doi{10.1007/s10817-009-9143-8}.

\bibitemdeclare{inproceedings}{vickers_geometric_1993}
\bibitem{vickers_geometric_1993}
\bibinfo{author}{Steven \surnamestart Vickers\surnameend}
  (\bibinfo{year}{1993}): \emph{\bibinfo{title}{Geometric {Logic} in {Computer}
  {Science}}}.
\newblock In: {\sl \bibinfo{booktitle}{Theory and {Formal} {Methods}}},
  \bibinfo{series}{Workshops in {Computing}}, \bibinfo{publisher}{Springer},
  pp. \bibinfo{pages}{37--54}, \doi{10.1007/978-1-4471-3503-6\_4}.

\bibitemdeclare{inproceedings}{wang_automated_2003}
\bibitem{wang_automated_2003}
\bibinfo{author}{Dongming \surnamestart Wang\surnameend}
  (\bibinfo{year}{2003}): \emph{\bibinfo{title}{Automated {Generation} of
  {Diagrams} with {Maple} and {Java}}}.
\newblock In \bibinfo{editor}{Michael \surnamestart Joswig\surnameend} \&
  \bibinfo{editor}{Nobuki \surnamestart Takayama\surnameend}, editors: {\sl
  \bibinfo{booktitle}{Algebra, {Geometry} and {Software} {Systems}}},
  \bibinfo{publisher}{Springer}, \bibinfo{address}{Berlin, Heidelberg}, pp.
  \bibinfo{pages}{277--287}, \doi{10.1007/978-3-662-05148-1\_15}.

\bibitemdeclare{inproceedings}{wilson_combining_2005}
\bibitem{wilson_combining_2005}
\bibinfo{author}{Sean \surnamestart Wilson\surnameend} \&
  \bibinfo{author}{Jacques~D. \surnamestart Fleuriot\surnameend}
  (\bibinfo{year}{2005}): \emph{\bibinfo{title}{Combining {Dynamic} {Geometry},
  {Automated} {Geometry} {Theorem} {Proving} and {Diagrammatic} {Proofs}}}.
\newblock In: {\sl \bibinfo{booktitle}{{ETAPS} {Satellite} {Workshop} on {User}
  {Interfaces} for {Theorem} {Provers} ({UITP})}},
  \bibinfo{publisher}{Springer}, \bibinfo{address}{Edinburgh}.

\bibitemdeclare{inproceedings}{winterstein_dr.doodle:_2004}
\bibitem{winterstein_dr.doodle:_2004}
\bibinfo{author}{Daniel \surnamestart Winterstein\surnameend}
  (\bibinfo{year}{2004}): \emph{\bibinfo{title}{Dr.{Doodle}: {A} {Diagrammatic}
  {Theorem} {Prover}}}.
\newblock In: {\sl \bibinfo{booktitle}{Proceedings of {IJCAR} 2004}},
  \doi{10.1007/978-3-540-25984-8\_24}.

\bibitemdeclare{article}{ye_visually_2009}
\bibitem{ye_visually_2009}
\bibinfo{author}{Zheng \surnamestart Ye\surnameend},
  \bibinfo{author}{Shang-Ching \surnamestart Chou\surnameend} \&
  \bibinfo{author}{Xiao-Shan \surnamestart Gao\surnameend}
  (\bibinfo{year}{2010}): \emph{\bibinfo{title}{Visually {Dynamic}
  {Presentation} of {Proofs} in {Plane} {Geometry}}}.
\newblock {\sl \bibinfo{journal}{Journal of Automated Reasoning}}
  \bibinfo{volume}{45}(\bibinfo{number}{3}), pp. \bibinfo{pages}{243--266},
  \doi{10.1007/s10817-009-9163-4}.

\bibitemdeclare{article}{ye_visually_2010}
\bibitem{ye_visually_2010}
\bibinfo{author}{Zheng \surnamestart Ye\surnameend},
  \bibinfo{author}{Shang-Ching \surnamestart Chou\surnameend} \&
  \bibinfo{author}{Xiao-Shan \surnamestart Gao\surnameend}
  (\bibinfo{year}{2010}): \emph{\bibinfo{title}{Visually {Dynamic}
  {Presentation} of {Proofs} in {Plane} {Geometry}, {Part} 1}}.
\newblock {\sl \bibinfo{journal}{J. Autom. Reason.}}
  \bibinfo{volume}{45}(\bibinfo{number}{3}), pp. \bibinfo{pages}{213--241},
  \doi{10.1007/s10817-009-9162-5}.

\bibitemdeclare{inproceedings}{ye_introduction_2011}
\bibitem{ye_introduction_2011}
\bibinfo{author}{Zheng \surnamestart Ye\surnameend},
  \bibinfo{author}{Shang-Ching \surnamestart Chou\surnameend} \&
  \bibinfo{author}{Xiao-Shan \surnamestart Gao\surnameend}
  (\bibinfo{year}{2011}): \emph{\bibinfo{title}{An {Introduction} to {Java}
  {Geometry} {Expert}}}.
\newblock In: {\sl \bibinfo{booktitle}{Post-proceedings of {Automated}
  {Deduction} in {Geometry} ({ADG} 2008)}}, {\sl \bibinfo{series}{Lecture
  {Notes} in {Computer} {Science}}} \bibinfo{volume}{6301},
  \bibinfo{publisher}{Springer-Verlag}, pp. \bibinfo{pages}{189--195},
  \doi{10.1007/978-3-642-21046-4\_10}.

\end{thebibliography}


\appendix

\section{Example: Varignon's theorem}
In this section, we provide as examples two different proofs of a statement in Euclidean geometry called Varignon's theorem.

\begin{theorem}
th\_varignon : $\forall A \; \forall B \; \forall C \; \forall D \; \forall I \; \forall J \; \forall K \; \forall L \; ( \neg col(B, D, A)\wedge \neg col(B, D, C)\wedge \neg col(A, C, B)\wedge \neg col(A, C, D)\wedge \neg col(I, J, K)\wedge B \neq D\wedge A \neq C\wedge midpoint(A, I, B)\wedge midpoint(B, J, C)\wedge midpoint(C, K, D)\wedge midpoint(A, L, D) \Rightarrow pG(I, J, K, L)\;)$
\end{theorem}

\setcounter{proofstepnum}{0}

{\scriptsize
\noindent{\em First Proof:}
\vspace{5pt}

\noindent Consider arbitrary $a$, $b$, $c$, $d$, $e$, $f$, $g$, $h$ such that:  $\neg col(b, d, a)$,  $\neg col(b, d, c)$,  $\neg col(a, c, b)$,  $\neg col(a, c, d)$,  $\neg col(e, f, g)$,  $b \neq d$,  $a \neq c$,  $midpoint(a, e, b)$,  $midpoint(b, f, c)$,  $midpoint(c, g, d)$,  $midpoint(a, h, d)$. It should be proved that $pG(e, f, g, h)$.
\vspace{5pt}

\proofstep{0}{$par(b, d, f, g)$ ({\scriptsize by MP, from $\neg col(b, d, c)$, $midpoint(c, g, d)$, $midpoint(b, f, c)$ using axiom triangle\_mid\_par\_strict; instantiation:  $A$ $\mapsto$  $b$,  $B$ $\mapsto$  $d$,  $C$ $\mapsto$  $c$,  $P$ $\mapsto$  $g$,  $Q$ $\mapsto$  $f$}) }
\proofstep{0}{$par(b, d, e, h)$ ({\scriptsize by MP, from $\neg col(b, d, a)$, $midpoint(a, h, d)$, $midpoint(a, e, b)$ using axiom triangle\_mid\_par\_strict; instantiation:  $A$ $\mapsto$  $b$,  $B$ $\mapsto$  $d$,  $C$ $\mapsto$  $a$,  $P$ $\mapsto$  $h$,  $Q$ $\mapsto$  $e$}) }
\proofstep{0}{$par(a, c, e, f)$ ({\scriptsize by MP, from $\neg col(a, c, b)$, $midpoint(b, f, c)$, $midpoint(a, e, b)$ using axiom triangle\_mid\_par\_strict; instantiation:  $A$ $\mapsto$  $a$,  $B$ $\mapsto$  $c$,  $C$ $\mapsto$  $b$,  $P$ $\mapsto$  $f$,  $Q$ $\mapsto$  $e$}) }
\proofstep{0}{$par(a, c, h, g)$ ({\scriptsize by MP, from $\neg col(a, c, d)$, $midpoint(c, g, d)$, $midpoint(a, h, d)$ using axiom triangle\_mid\_par\_strict; instantiation:  $A$ $\mapsto$  $a$,  $B$ $\mapsto$  $c$,  $C$ $\mapsto$  $d$,  $P$ $\mapsto$  $g$,  $Q$ $\mapsto$  $h$}) }
\proofstep{0}{$par(e, f, g, h)$ ({\scriptsize by MP, from $par(a, c, e, f)$, $par(a, c, h, g)$, $\neg col(e, f, g)$ using axiom lemma\_par\_\-trans; instantiation:  $A$ $\mapsto$  $e$,  $B$ $\mapsto$  $f$,  $C$ $\mapsto$  $a$,  $D$ $\mapsto$  $c$,  $E$ $\mapsto$  $g$,  $F$ $\mapsto$  $h$}) }
\proofstep{0}{$par(f, g, e, h)$ ({\scriptsize by MP, from $par(b, d, f, g)$, $par(b, d, e, h)$, $par(e, f, g, h)$ using axiom lemma\_par\_\-trans; instantiation:  $A$ $\mapsto$  $f$,  $B$ $\mapsto$  $g$,  $C$ $\mapsto$  $d$,  $D$ $\mapsto$  $b$,  $E$ $\mapsto$  $e$,  $F$ $\mapsto$  $h$}) }
\proofstep{0}{$pG(e, f, g, h)$ ({\scriptsize by MP, from $par(e, f, g, h)$, $par(e, f, g, h)$, $par(f, g, e, h)$ using axiom lemma\_par2\_\-pg; instantiation:  $A$ $\mapsto$  $e$,  $B$ $\mapsto$  $f$,  $C$ $\mapsto$  $g$,  $D$ $\mapsto$  $h$}) }
\proofstep{0}{Proved by assumption! ({\scriptsize by QEDas})}
}

\begin{figure}
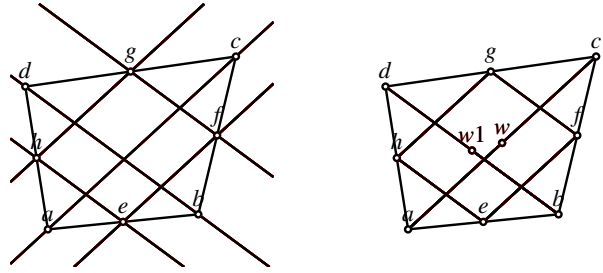

\begin{center}
\input{PROOFvarignonpQF_BV_illustration.tkz}	  
\hspace*{1cm}
\input{PROOFvarignon2pQF_BV_illustration.tkz}	  
\end{center}
\caption{Illustrations for two proofs of Varignon's theorem}
\end{figure}

\setcounter{proofstepnum}{0}

{\scriptsize
\noindent{\em Second Proof:}
\vspace{5pt}

\noindent Consider arbitrary $a$, $b$, $c$, $d$, $e$, $f$, $g$, $h$ such that:  $f \neq h$,  $b \neq d$,  $d \neq b$,  $a \neq c$,  $c \neq a$,  $midpoint(a, e, b)$,  $midpoint(b, f, c)$,  $midpoint(c, g, d)$,  $midpoint(a, h, d)$. It should be proved that $pG(e, f, g, h)$.
\vspace{5pt}

\proofstep{0}{Let $w$ be such that $midpoint(a, w, c)$ ({\scriptsize by MP, from $a \neq c$ using axiom lemma\_midpoint\_existence; instantiation:  $A$ $\mapsto$  $a$,  $B$ $\mapsto$  $c$}) }
\proofstep{0}{Let $w1$ be such that $midpoint(d, w1, b)$ ({\scriptsize by MP, from $d \neq b$ using axiom lemma\_midpoint\_existence; instantiation:  $A$ $\mapsto$  $d$,  $B$ $\mapsto$  $b$}) }
\proofstep{0}{$cong(a, w, g, h)$ ({\scriptsize by MP, from $c \neq a$, $midpoint(a, w, c)$, $midpoint(c, g, d)$, $midpoint(a, h, d)$ using axiom lemma\_triangle\_mid\_par\_cong\_1; instantiation:  $A$ $\mapsto$  $d$,  $B$ $\mapsto$  $a$,  $C$ $\mapsto$  $c$,  $P$ $\mapsto$  $w$,  $Q$ $\mapsto$  $g$,  $R$ $\mapsto$  $h$}) }
\proofstep{0}{$cong(a, w, f, e)$ ({\scriptsize by MP, from $c \neq a$, $midpoint(a, w, c)$, $midpoint(b, f, c)$, $midpoint(a, e, b)$ using axiom lemma\_triangle\_mid\_par\_cong\_1; instantiation:  $A$ $\mapsto$  $b$,  $B$ $\mapsto$  $a$,  $C$ $\mapsto$  $c$,  $P$ $\mapsto$  $w$,  $Q$ $\mapsto$  $f$,  $R$ $\mapsto$  $e$}) }
\proofstep{0}{$tP(c, a, h, g)$ ({\scriptsize by MP, from $a \neq c$, $midpoint(a, w, c)$, $midpoint(a, h, d)$, $midpoint(c, g, d)$ using axiom lemma\_triangle\_mid\_par\_cong\_1; instantiation:  $A$ $\mapsto$  $d$,  $B$ $\mapsto$  $c$,  $C$ $\mapsto$  $a$,  $P$ $\mapsto$  $w$,  $Q$ $\mapsto$  $h$,  $R$ $\mapsto$  $g$}) }
\proofstep{0}{$tP(b, d, h, e)$ ({\scriptsize by MP, from $d \neq b$, $midpoint(d, w1, b)$, $midpoint(a, h, d)$, $midpoint(a, e, b)$ using axiom lemma\_triangle\_mid\_par\_cong\_1; instantiation:  $A$ $\mapsto$  $a$,  $B$ $\mapsto$  $b$,  $C$ $\mapsto$  $d$,  $P$ $\mapsto$  $w1$,  $Q$ $\mapsto$  $h$,  $R$ $\mapsto$  $e$}) }
\proofstep{0}{$cong(b, w1, h, e)$ ({\scriptsize by MP, from $d \neq b$, $midpoint(d, w1, b)$, $midpoint(a, h, d)$, $midpoint(a, e, b)$ using axiom lemma\_triangle\_mid\_par\_cong\_1; instantiation:  $A$ $\mapsto$  $a$,  $B$ $\mapsto$  $b$,  $C$ $\mapsto$  $d$,  $P$ $\mapsto$  $w1$,  $Q$ $\mapsto$  $h$,  $R$ $\mapsto$  $e$}) }
\proofstep{0}{$cong(h, e, h, e)$ ({\scriptsize by MP, from $cong(b, w1, h, e)$, $cong(b, w1, h, e)$ using axiom lemma\_congruence\-transitive; instantiation:  $A$ $\mapsto$  $h$,  $B$ $\mapsto$  $e$,  $C$ $\mapsto$  $b$,  $D$ $\mapsto$  $w1$,  $E$ $\mapsto$  $h$,  $F$ $\mapsto$  $e$}) }
\proofstep{0}{$cong(g, h, e, f)$ ({\scriptsize by MP, from $cong(a, w, g, h)$, $cong(a, w, f, e)$ using axiom lemma\_congruence\-transitive; instantiation:  $A$ $\mapsto$  $g$,  $B$ $\mapsto$  $h$,  $C$ $\mapsto$  $w$,  $D$ $\mapsto$  $a$,  $E$ $\mapsto$  $e$,  $F$ $\mapsto$  $f$}) }
\proofstep{0}{$cong(b, w1, g, f)$ ({\scriptsize by MP, from $d \neq b$, $midpoint(d, w1, b)$, $midpoint(c, g, d)$, $midpoint(b, f, c)$ using axiom lemma\_triangle\_mid\_par\_cong\_1; instantiation:  $A$ $\mapsto$  $c$,  $B$ $\mapsto$  $b$,  $C$ $\mapsto$  $d$,  $P$ $\mapsto$  $w1$,  $Q$ $\mapsto$  $g$,  $R$ $\mapsto$  $f$}) }
\proofstep{0}{$cong(f, g, e, h)$ ({\scriptsize by MP, from $cong(b, w1, g, f)$, $cong(b, w1, h, e)$ using axiom lemma\_congruence\-transitive; instantiation:  $A$ $\mapsto$  $f$,  $B$ $\mapsto$  $g$,  $C$ $\mapsto$  $b$,  $D$ $\mapsto$  $w1$,  $E$ $\mapsto$  $e$,  $F$ $\mapsto$  $h$}) }
\proofstep{0}{$tP(b, d, g, f)$ ({\scriptsize by MP, from $d \neq b$, $midpoint(d, w1, b)$, $midpoint(c, g, d)$, $midpoint(b, f, c)$ using axiom lemma\_triangle\_mid\_par\_cong\_1; instantiation:  $A$ $\mapsto$  $c$,  $B$ $\mapsto$  $b$,  $C$ $\mapsto$  $d$,  $P$ $\mapsto$  $w1$,  $Q$ $\mapsto$  $g$,  $R$ $\mapsto$  $f$}) }
\proofstep{0}{$tP(a, c, f, e)$ ({\scriptsize by MP, from $c \neq a$, $midpoint(a, w, c)$, $midpoint(b, f, c)$, $midpoint(a, e, b)$ using axiom lemma\_triangle\_mid\_par\_cong\_1; instantiation:  $A$ $\mapsto$  $b$,  $B$ $\mapsto$  $a$,  $C$ $\mapsto$  $c$,  $P$ $\mapsto$  $w$,  $Q$ $\mapsto$  $f$,  $R$ $\mapsto$  $e$}) }
\proofstep{0}{$tP(g, h, f, e)$ ({\scriptsize by MP, from $tP(c, a, h, g)$, $tP(a, c, f, e)$ using axiom lemma\_tP\_trans; instantiation:  $A$ $\mapsto$  $g$,  $B$ $\mapsto$  $h$,  $C$ $\mapsto$  $a$,  $D$ $\mapsto$  $c$,  $E$ $\mapsto$  $f$,  $F$ $\mapsto$  $e$}) }
\proofstep{0}{$tP(f, g, e, h)$ ({\scriptsize by MP, from $tP(b, d, g, f)$, $tP(b, d, h, e)$ using axiom lemma\_tP\_trans; instantiation:  $A$ $\mapsto$  $f$,  $B$ $\mapsto$  $g$,  $C$ $\mapsto$  $b$,  $D$ $\mapsto$  $d$,  $E$ $\mapsto$  $e$,  $F$ $\mapsto$  $h$}) }
\proofstep{0}{$pG(e, f, g, h)$ ({\scriptsize by MP, from $f \neq h$, $cong(g, h, e, f)$, $cong(f, g, e, h)$, $tP(f, g, e, h)$, $tP(g, h, f, e)$ using axiom lemma\_par\_par\_cong\_cong\_para\-llelogram; instantiation:  $A$ $\mapsto$  $e$,  $B$ $\mapsto$  $f$,  $C$ $\mapsto$  $g$,  $D$ $\mapsto$  $h$}) }
\proofstep{0}{Proved by assumption! ({\scriptsize by QEDas})}
}

\end{document}